\documentclass[twoside]{article}
\usepackage{fleqn,espcrc2}

\usepackage{graphicx}
\usepackage{epsfig}
\usepackage[figuresright]{rotating}

\newcommand{\Sign}{{\rm Phase}}
\newcommand{\AmS}{{\protect\the\textfont2
  A\kern-.1667em\lower.5ex\hbox{M}\kern-.125emS}}

\title{QCD at a Finite Density of Static Quarks}

\author{Shailesh Chandrasekharan\address[DUKE]{Department of Physics, 
        Duke University, P.O. Box 90305, Durham NC 27708-0305, USA}%
        \thanks{Part of this work was done in collaboration with 
		M.Alford, J. Cox, and U.-J. Wiese and was supported by
		a US Department of Energy grant \# DE-FG02-96ER40945}}
\begin{document}

\begin{abstract}
Recently, cluster methods have been used to solve a variety of sign 
problems including those that arise in the presence of fermions.
In all cases an analytic partial re-summation over a class of
configurations in the path integral was necessary. Here the new ideas 
are illustrated using the example of QCD at a finite density of static 
quarks. In this limit the sign problem simplifies since the fermionic 
part decouples. Furthermore, the problem can be solved completely when 
the gauge dynamics is replaced by a Potts model. On the other hand in 
QCD with light quarks the solution will require a partial re-summation 
over both fermionic and gauge degrees of freedom. The new approach points 
to unexplored directions in the search for a solution to this more 
challenging sign problem.
\vspace{1pc}
\end{abstract}

\maketitle

\section{INTRODUCTION}

Although lattice field theories allow for a completely non-perturbative 
definition of a quantum field theory, computational limitations impose 
severe restrictions on our ability to perform calculations in them. At 
the beginning of the new millennium many important questions that
arise in a variety of fields like strongly interacting fermionic systems, 
frustrated systems, systems with $\theta$-vacuum, chiral gauge theories 
and real time dynamics in quantum field theories, remain unanswered.
A common problem that hinders progress in all these areas is the ``sign 
problem'', which arises due to the inability to isolate the appropriate 
probabilistic measure for a stochastic computation.

One is typically interested in the partition function of a quantum
many body system which can be compactly written as
\begin{equation}
Z \;=\; {\rm Tr}\left[ \exp(-H/T)\right] \;=\; \sum_{[C]} W[C]
\end{equation}
where the sum over configurations $C$ is used to evaluate the
trace through the Feynman path integral in a convenient basis.
In general $W[C]$, the Boltzmann weight of the configuration,
is not guaranteed to be positive definite. The difficulty in finding 
the right class of configurations such that $W(C)$ is positive definite 
is referred to as the sign problem. However, since $Z$ itself
is positive definite a summation over a sufficiently large number 
of configurations should always yield a positive answer. Unfortunately, 
the number of configurations required grows exponentially with the 
volume; it is hopeless to require even a computer to perform the 
re-summation.

Recently, a lot of progress has been made in solving sign problems 
using cluster methods since it is sometimes possible to analytically 
sum over all the configurations that are obtained through cluster 
flips. Furthermore, this class of configurations can grow 
exponentially with volume. Interestingly, every known type of cluster 
algorithm has been applied to solve to a new type of sign problem. 
The first solution was found in a classical $O(3)$ model with a $\theta$ 
parameter which makes the action complex \cite{Bie95}. It was shown that 
a summation over flips of Wolff clusters \cite{Wol89} produced only 
positive (or
zero) weights. More recently, it was discovered that in certain four-Fermi
models a similar result emerges when one uses loop clusters known 
from quantum spin systems \cite{Eve93}. This has been used to eliminate 
sign problems that arise due to Fermi statistics \cite{Cha99a,Cha99b}. 
Sign problems in Potts models due to a complex action can be solved 
using clusters of the Swendsen-Wang algorithm\cite{Swe87}. 

In this article the above progress is reviewed using the example 
of the Potts model which relevant to QCD at a finite density of static 
quarks. The next section contains a review of the sign problem in QCD 
with a new perspective on its ``physical origin''. It is argued that the 
sign problem has two parts, one coming from Gauss law type constraints 
and the other from fermion permutations. The later sections show that
the recent progress has produced examples in which both types of sign 
problems can be solved separately. In simple cases it may be possible 
to find solutions to the combined problem in the near future.
 
\section{THE SIGN PROBLEM IN QCD}

The partition function in QCD is usually written as a
path integral over fermionic and gauge fields. Instead
of representing fermions as Grassmann variables, for 
computational reasons they are integrated out in favor 
of a fermion determinant. The final result
takes the form
\begin{equation}
Z_{\rm QCD} \;=\; \int\;[dU]\;\exp\left(-S_G[U]\right)
\;{\rm Det}(D[U])
\label{zqcd}
\end{equation}
where $D$ is the Dirac operator. Given that the gauge field 
action $S_G[U]$ is real the usual claim is that sign problems 
arise from the fermion determinant. As is well known in the 
presence of a chemical potential the fermion determinant 
${\rm Det}(D[U])$ becomes complex. Although this description 
is quite accurate, it is a bit unsatisfactory since a lot of 
interesting dynamics of quarks has been pushed into the 
calculation of the fermion determinant. Exposing the anatomy 
of the fermion determinant may reveal some deeper 
``physical origin'' to the sign problem. This can be 
illustrated by a simple example.

Consider compact QED instead of QCD. The partition function 
in that case is similar to (\ref{zqcd}) except that the 
integration is over $U(1)$ gauge fields instead of $SU(3)$. 
Again adding a chemical potential makes the fermion determinant
complex and hence introduces a sign problem for the same reasons 
as in QCD. But in this case it is possible to understand the 
origin of the sign problem at a more deeper level. If the fermion 
determinant is expanded as a sum over electron and positron world 
line configurations, the configurations that dominate at non-zero 
chemical potentials would contain more electrons than positrons.
However, due to Gauss's law it is impossible to 
have more electrons than positrons in a periodic box. Hence,
the weight of such configurations must be zero once Gauss's
law is imposed. Since it is the integral over the gauge fields 
that enforces Gauss's law, it is natural to have both positive 
and negative contributions in the path integral for a fixed gauge 
field configuration. The negative signs can help cancel the 
wrong configurations. Understanding the ``physical'' origin of 
the sign problem makes the solution clear. In the case of QED, 
one must throw away all fermionic configurations in which there 
are more electrons than positrons. This effectively means the 
chemical potential has no effect on the partition function which 
is a well known fact \cite{Ben92}.

Unfortunately, fermionic world line configurations have sign
problems of their own that arise from the Pauli principle. When 
fermions travel in imaginary time their positions can get permuted. 
Whenever this permutation is odd the configuration weight is negative. 
Thus, performing the integration over the gauge field configurations 
alone in general will not yield a positive weight for a fermionic 
configuration. In fact the sign problem in QCD arises due to an
intricate mixture of fermion permutation sign and a more difficult
constraints one of which is the Gauss's law. The example from QED 
suggests that it may be useful to consider a regrouping of both 
fermion and gauge field configurations in the search to
find positive definite weights. 

\section{THE STATIC QUARK LIMIT}

There are special limits where the sign problem becomes much simpler.
One is the limit of static quarks. Here the fermions cannot permute
with each other and so problems arise only from Gauss's law type
constraints. This simplification has prompted a variety of research 
on the physics of QCD with a finite density of static quarks. However,
adding chemical potential to quarks that are infinitely massive
is a bit subtle. In order to induce a finite density of static quarks 
one needs to take the chemical potential to infinity along with the 
quark mass always keeping the chemical potential larger than the rest 
mass of the quarks. This limit was originally formulated in \cite{Ben92} 
and later studied in \cite{Blu96}. More recently the partition function 
for a fixed baryon number in the static quark limit was studied in
\cite{Eng99}. An equivalent but simpler approach to the limit of static 
quarks can be formulated by recognizing that 
the partition function for $n$ quarks and $\bar{n}$ antiquarks of mass $m$ in 
the static limit is given by
\begin{equation}
Z_{n,\bar{n}} \;=\;
\int\;[dU]\;
\frac{\Phi^n}{n!}\;\frac{{\Phi^*}^{\bar{n}}}{\bar{n}!}\;
{\rm e}^{-S_G[U] - m(n+\bar{n})N_t}
\end{equation}
where $S_G[U]$ is the gauge action on a $N_s^3 N_t$ lattice. The gauge
coupling $\beta$ is implicit in the definition of $S_G$. The quantity 
$\Phi$ is the sum over the trace of Polyakov lines at every spatial 
point ${\mathbf x}$,
\begin{equation}
\Phi = \sum_{\mathbf x} Tr\left(\prod_{t=0}^{N_t-1} U_0({\mathbf x},t)\right),
\end{equation}
with $U_0({\mathbf x},t)$ representing the time like links. The grand 
canonical partition function is then given by
\begin{eqnarray}
Z &=& \sum_{n,\bar{n}} Z_{n,\bar{n}} \exp\left(\mu(n-\bar{n})/T\right) \\
&=& \int\;[dU]\;
{\rm e}^{-S_G[U] + h\;\Phi + h^\prime\;\Phi^*}
\label{zsqcd}
\end{eqnarray} 
where $h={\rm e}^{-(m-\mu)N_t}$ and $h^\prime = {\rm e}^{-(m+\mu)N_t}$.
The static quark limit is obtained by taking the $m\rightarrow \infty$ 
and $\mu\rightarrow \infty$ limits with $h={\rm e}^{-(m-\mu)N_t}$ fixed.
In this limit $h^\prime \rightarrow 0$. Note that although the above 
approach treats quarks as bosons instead of fermions, at finite densities 
in the continuum limit it is impossible to distinguish between them.

  Due to the complex nature of ${\rm e}^{h\Phi}$ the partition function 
of eq. (\ref{zsqcd}) suffers from a sign problem albeit a much simpler 
one as compared to eq. (\ref{zqcd}). Absorbing the magnitude of this term 
in the Boltzmann weight, one can define the average of the phase as,
\begin{equation}
\langle\Sign\rangle \;=\;
\frac{\int\;[dU]\;{\rm e}^{-S_G[U] + h{\rm Re}\Phi}\;
{\rm e}^{ih{\rm Im}\Phi}}
{\int\;[dU]\;{\rm e}^{-S_G[U] + h{\rm Re}\Phi}}.
\end{equation} 
Since $\langle\Sign\rangle$ is a ratio of two partition functions, in 
the large volume limit it will vanish exponentially. In order to make 
progress without solving the sign problem it is necessary to remain in 
volumes where the average sign is not too small. This usually puts severe 
restrictions on our ability to address interesting questions regarding 
the phase transition. 

\begin{figure}
\begin{center}
\vskip-0.3in
\includegraphics[width=0.45\textwidth]{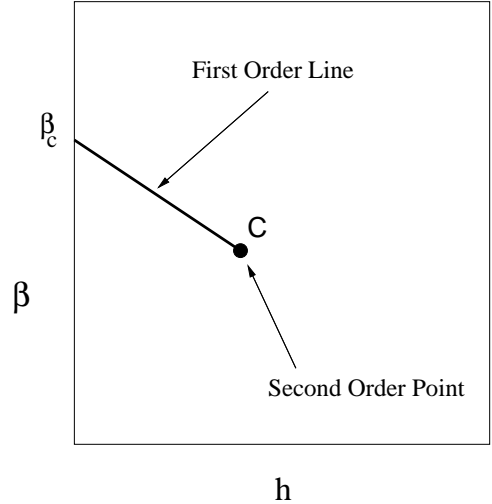}
\end{center}
\vskip-0.7in
\caption[]{The phase diagram of static QCD. The point $C$ is expected
to be in the universality class of the 3-d Ising model.}
\label{pdiag}
\end{figure}

  The phase diagram in the $\beta-h$ plane was first studied in 
\cite{Blu96}. Figure \ref{pdiag} shows the most likely scenario 
based on the well known fact that there is a first order transition on 
the $h=0$ line at $\beta_c$ due to the spontaneous breaking of the
${\mathbf{Z}_3}$ center symmetry. When $h\neq 0$ the symmetry is
explicitly broken. However, the first order phase transition should 
typically 
persist for non-zero $h$ with decreasing latent heat ending at the 
second order critical point $C$ which should be governed by the 3-d
Ising universality class. Although, the study in \cite{Blu96} could not 
confirm this picture it showed that the point $C$ must be quite close
to the $h=0$ axis. In particular the well known discontinuity in the 
Polyakov line at $\beta_c$ disappeared at the smallest $h$ studied. 
This information suggests that with out solving the sign problem it 
may still be possible to study the behavior of the model close to 
the critical line of the phase diagram on lattices whose size is 
constrained by the closeness of the point $C$ to the $h=0$ axis.

\section{SOLVING THE SIGN PROBLEM} 

  An interesting challenge is to be able to solve the sign problem
arising in eq. (\ref{zsqcd}). Such a solution can perhaps teach us 
about solutions to other more interesting sign problems in gauge
theories. In order to make progress it is useful to learn from a 
simpler example where the gauge dynamics is replaced with a simple 
3-state Potts model. The new partition function takes the form of
\begin{equation}
Z = \sum_{z} {\rm e}^{\;\beta\;E[z] + h M[z] + h^\prime M^*[z]}
\label{zpotts}
\end{equation}
with the energy and magnetization defined by
\begin{equation}
E[z] = \sum_{{\mathbf x},i} \delta_{z_{\mathbf x},z_{{\mathbf x}+\hat{i}}},\;\;
M[z] \;=\;\sum_{\mathbf x} z_{\mathbf x}.
\end{equation}
Due to its symmetry properties, the three state Potts model with 
$z_{\mathbf x} \in \{1, {\rm e}^{i2\pi/3} , {\rm e}^{-i2\pi/3}\}$, has 
been considered a useful effective model for static QCD close to the phase 
transition \cite{Yaf82,DeG83}.

  An interesting solution to the sign problem in (\ref{zpotts}) was 
proposed in \cite{Con00}. It was shown that the partition function
can be rewritten in terms of new variables which describe classical 
statistical mechanics arising from the Hamiltonian,
\begin{equation}
H = \sum_{{\mathbf x},i} \sigma|\;l_{{\mathbf x},i}|\;
+\sum_{\mathbf x} (M\;|\;n_{\mathbf x}| - \mu \;n_{\mathbf x}),
\end{equation}
that represents the energy of configurations labeled by quark number 
distribution $n_{\mathbf x} \in \{-3,-2,...,3\}$ associated with
sites and color flux variables $l_{{\mathbf x},i}\in\{-1,0,1\}$ associated
with links. The configurations further obey the Gauss's law constraint
\begin{equation}
\sum_{i=1}^3(l_{{\mathbf x},i}-l_{{\mathbf x},-i}) = 
n_{\mathbf x}\; \mbox{mod} \;\;3.
\end{equation}
This type of flux tube models is well known in the literature \cite{Pat84}.
Since the Hamiltonian is real, except for the complications arising
from the Gauss law constraint, the sign problem is completely solved.
The introduction of new variables has helped in re-summing over
classes of configurations in the original variables. In \cite{Con00}
a metropolis algorithm was used to study the flux tube model.

 There is another way to solve the sign problem arising in (\ref{zpotts})
based on cluster algorithms for Potts models \cite{Swe87}. The essential
idea in this case is to write the weight associated with a bond, 
\begin{equation}
{\rm e}^{\kappa \delta_{z_x,z_y}}\;=\;\sum_{b=0,1}
\left[\delta_{b,1}\delta_{z_x,z_y}({\rm e}^\kappa-1) + \delta_{b,0}\right],
\end{equation}
as a sum over new bond variables $b$, with $b=1$ representing the
presence of a connection between the sites and $b=0$ representing
the absence of a connection. Every configuration $[z,b]$ of bonds 
and Potts spins is made up of a connected cluster $({\cal C})$ of sites, 
with the property that each connected cluster of sites always contains 
the same Potts spin $z_{\cal C}$. This property allows the Boltzmann 
weight of each configuration $[z,b]$ to be written as a product over 
cluster weights and the partition function takes the form
\begin{equation}
Z = \sum_{[z,b]} \;\left\{\prod_{{\cal C}} W({\cal C})\;
\exp\left( L_{\cal C}\;[h z_{\cal C} + h^\prime z^*_{\cal C}]\right)\right\},
\end{equation}
where $L_{\cal C}$ represents the size of the cluster ${\cal C}$ and
$W({\cal C})=({\rm e}^\kappa-1)^{B_{\cal C}}$ where $B_{\cal C}$ stands
for the number of $b=1$ bonds in the cluster. Since $W({\cal C})$ does
not depend on the spin variables, it is easy to perform a sum over the
allowed cluster spin variables $z_{\cal C}\in\{1,{\rm e}^{i2\pi/3}$ and 
${\rm e}^{-i2\pi/3}\}$. The partition function can finally be written as
\begin{equation}
Z = \sum_{[b]} \;\left\{\prod_{{\cal C}} W({\cal C})\;
\left(\sum_{z} {\rm e}^{L_{\cal C}\;[h z + h^\prime z^*]}\right)\right\}.
\label{zpottsc}
\end{equation}
It is easy to check that the Boltzmann weight in (\ref{zpottsc}) for
$h,h^\prime > 0$ is always positive which solves the sign problem.

    The two solutions to the sign problem sketched above suggest that 
an appropriate regrouping of the gauge configurations may yield a
solution in the case of QCD with static quarks. However, this has not 
yet been achieved and remains an open problem for the future. 
On the other hand, based on the strong coupling expansion, a solution of 
the type found in the flux representation of the Potts model may be easy
to find. A cluster type solution, although more exciting, could be difficult.

\section{RESULTS IN THE POTTS MODEL}

Recently, the partition function of eq. (\ref{zpotts}) has been
studied numerically for $h=h^\prime$. Since in this case the action is 
real the model does not suffer from a sign problem. The critical 
region has been analyzed in great detail and the critical end point 
$C$ is found to be $(\beta_c=0.54938(2),h_c=0.000516(7))$ 
\cite{Kar00,Kar00a}. The reader is referred to the original article 
for further results and details. 

\begin{figure}[t]
\begin{center}
\vskip-0.3in
\includegraphics[width=0.46\textwidth,height=0.42\textwidth]{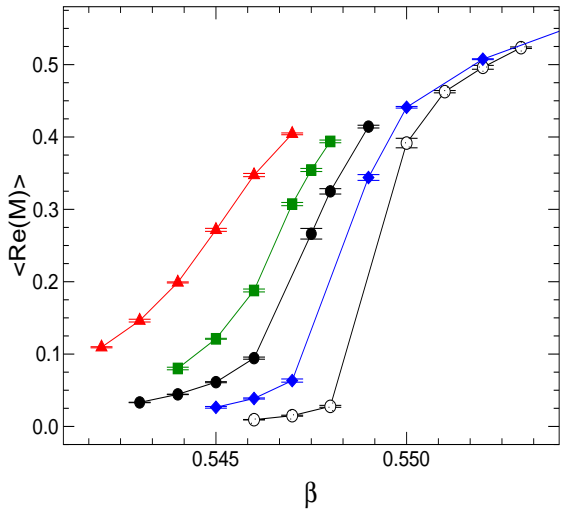}
\end{center}
\vskip-0.5in
\caption[]{Expectation value  of ${\rm Re}(M)$ as a function of $\beta$ for 
$h=0.0005$ (open circles), $0.001$ (diamonds), $0.0015$ (filled circles),
$0.002$ (squares), and $0.0025$ (triangles) in the infinite volume limit.}
\label{poly}
\end{figure}

The model with $h^\prime=0$ is more closer in spirit to the physics 
of a finite density of static quarks. Although in this case the action 
is complex, the sign problem can be completely solved as discussed
in the previous section. Using this solution and an algorithm described
in \cite{Alf00}, the behavior of the magnetization $M$ was studied on 
lattices as large as $40^3$. Using finite size scaling analysis the 
were extrapolated to the infinite volume limit. In figure \ref{poly} 
some results are plotted as a function of $\beta$ for various values of 
$h$. Clearly, even for $h=0.002$ the transition has already turned into 
a smooth cross over. This result is quite consistent with the results 
of \cite{Blu96}.

\begin{figure}[t]
\begin{center}
\vskip-0.3in
\includegraphics[width=0.46\textwidth,height=0.42\textwidth]{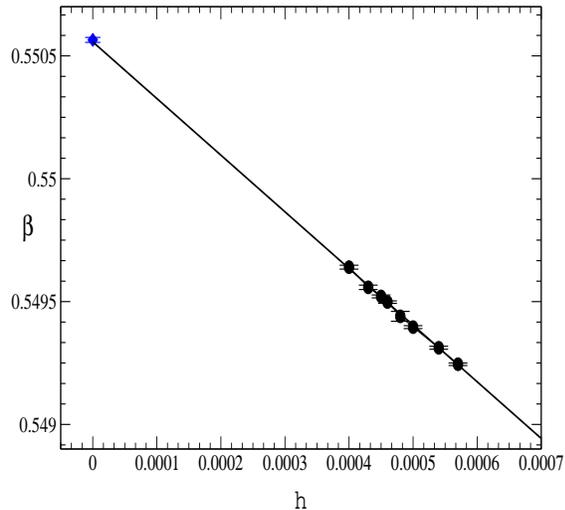}
\end{center}
\vskip-0.5in
\caption[]{The critical line in the $\beta-h$ plane. The data
points represent the infinite volume extrapolations of the peaks
of the specific heat. The solid line is a plot of eq. (\ref{cline}).}
\label{pdiag2}
\end{figure}

Since the critical line is very close to the $h=0$ axis, the sign problem 
is rather mild. An estimate shows that 
$\langle \Sign\rangle \simeq \exp(-N_s^3/73^3)$ suggesting that only for 
$N_s>80$ will the solution to the sign problem become necessary. In fact 
it appears to be more efficient to simulate the theory with the real 
action and absorb the complex phase into observables for smaller lattice
sizes. Using this technique simulations were extended to $80^3$ lattices 
and the location of the critical line was determined using peaks in 
the specific heat. Figure \ref{pdiag2}, shows the data along with a fit 
to the critical line
\begin{equation}
\beta = 0.550557(21) - h/0.4337(78).
\label{cline}
\end{equation}
This line extrapolates very nicely to the first order critical point 
on the $h=0$ axis, which is well known to be $\beta_c = 0.550565(10)$ 
\cite{Jan97}. A preliminary analysis of the critical behavior near
the end point $C$ shows consistency with the 3-d Ising model. The
critical point is found to be shifted mildly to 
$(\beta_c\sim 0.5495(1),h_c\sim 0.000468(5))$. A more detailed 
discussion of the analysis will be given in \cite{Alf00}.

\section{FERMION SIGN PROBLEMS}

The difficulties associated with finite density QCD are shared by much
simpler four-Fermi models like the repulsive Hubbard model away from 
half filling. The conventional approach in interacting fermion models 
is to rewrite the interactions as fermion bilinears using  
auxiliary fields so that the fermions can be integrated out
in favor of a fermion determinant, very much like in QCD. Hence it
is not surprising that similar problems arise even in simple cases.
An interesting question is whether one can deal with fermion interactions 
differently. Perhaps a more direct approach involving fermionic 
configurations will yield novel solutions if the fermion sign problem
is tackled.

Some years ago this approach was investigated in \cite{Kar89}. Starting 
with finite density QCD in the strong coupling limit the gauge integration 
was performed explicitly and a purely fermionic action was obtained. 
The fermionic theory had six-Fermi couplings representing a baryonic
mass term. Although one would naively disregard this approach as 
difficult, the fermionic model could be solved with a novel algorithm 
which used a clever partial solution to the sign problem in the model. 
Unfortunately, the physics of the strong coupling limit appeared
rather uninteresting at high densities. It would be interesting to 
investigate this approach further and ask if one can extend the study 
beyond the leading strong coupling approximation.

\begin{figure}[t]
\begin{center}
\vskip-0.3in
\includegraphics[width=0.45\textwidth]{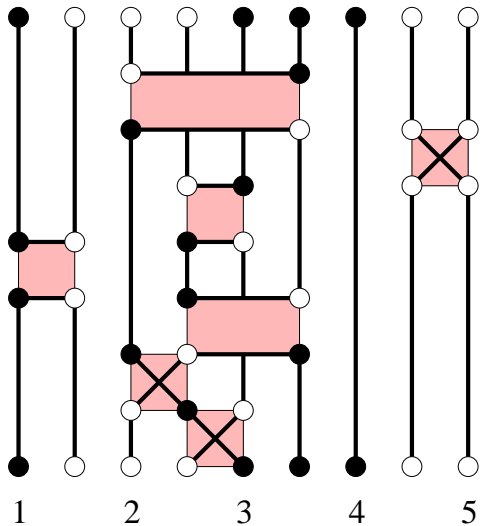}
\end{center}
\vskip-0.7in
\caption[]{A typical configuration of fermion occupation numbers
and clusters. The filled and open circles represent $n=1,0$.
The dark lines represent connected bonds. Shaded regions represent 
interactions. There are five clusters in the configuration.}
\label{fconf}
\end{figure}

Over the last couple of years a new class of solutions to sign problems
in a variety of four-Fermi models have emerged \cite{Cha99a,Cha99b}. 
The models are formulated directly in the fermionic Fock space, and 
configurations are labeled by fermion world lines. When the fermion 
permutation sign is taken into account with other local negative signs 
arising from transfer matrix elements, the partition function can 
be written as
\begin{equation}
Z = \sum_{[n]} {\rm Sign}[n]\;W[n]
\label{zferm}
\end{equation}
where $[n]$ represents a fermion occupation number configuration. Since
the Hilbert space of fermion occupation number states is isomorphic to
that of a spin-1/2 particle, the cluster algorithms for quantum spin 
systems can be used to formulate algorithms for the fermions \cite{Wie93}. 
Except for subtleties arising due to Fermi statistics, this essentially 
means that one is able to rewrite the partition function (\ref{zferm}) in 
terms of fermion occupation numbers and bonds $b$,
\begin{equation}
Z = \sum_{[n,b]} {\rm Sign}[n,b]\;W[n,b],
\label{zfermc}
\end{equation}
which is analogous to the example in the Potts model discussed in
section 4. Figure \ref{fconf} shows a typical fermionic configuration 
with an odd permutation along with clusters which have the property 
that by changing all the fermion occupation numbers within a cluster 
from 0 to 1 and vice-versa, one obtains another allowed configuration. 
This update is referred to as a cluster flip.

\begin{figure}
\begin{center}
%\vskip-0.3in
\includegraphics[width=0.45\textwidth]{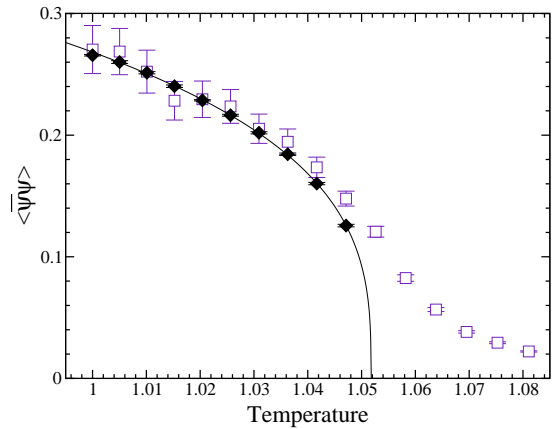}
\end{center}
\vskip-0.4in
\caption[]{The critical behavior of the chiral condensate as a function 
of temperature in a four-Fermi model. The open squares represent results
from a meron algorithm on a $32^3$ lattice at a mass of $m=0.001$. The 
black circles represent the infinite volume extrapolation. The complete 
results are presented in \cite{Cha00b}.}
\label{psi}
\end{figure}

Recently, a relation between the change in the fermion permutation
sign due to a cluster flip and the cluster topology was discovered 
\cite{Cha00a}. The relation showed that in certain models the 
sum over Boltzmann weights of configurations obtained through 
cluster flips for a fixed set of clusters is always positive; a 
result strikingly similar to the example of the Potts model. This
observation has been used to design an efficient fermion cluster 
algorithm and is referred to as the ``meron cluster algorithm''. It 
has been applied extensively to the study of the critical behavior 
near a ${\mathbf Z}_2$ finite temperature chiral phase transition with 
staggered fermions \cite{Cha99c,Cha00b}. This study for the first time 
demonstrated the emergence of Ising critical behavior in a fermionic model 
confirming dimensional reduction. Figure \ref{psi} shows the chiral 
condensate as a function of temperature in the staggered fermion mode. 
The scaling form $A(T_c-T)^{0.314(7)}$ consistently describes the data 
which is with in two standard deviations of the behavior in the 
3-d Ising model \cite{Cha00b}. Dimensional reduction in fermionic models
have been questioned in the past based on large $N$ calculations and 
results from hybrid Monte Carlo algorithms \cite{Kog95}. The
studies using meron algorithms are beginning to convincingly show the
validity of the conventional picture.

  Although the number of models that can be solved with a meron cluster 
algorithm is limited, it continues to grow with time. Recently, 
new models in the family of the attractive and repulsive Hubbard models
were found to be solvable. It is possible to add a chemical potential
in the attractive model without violating the properties necessary 
for the meron algorithm to work. Such models are solvable with the hybrid 
Monte Carlo algorithm. However, previous studies were limited to small
sizes \cite{Lac96}. The meron algorithm on the other hand can be implemented 
on much larger lattices with relative ease. Recently, this has allowed a high 
precision study of the Kosterlitz-Thouless transition. More details of 
this work can be found in \cite{Osb00}.

\section{FUTURE DIRECTIONS}

The amount of progress over the last couple of years has been remarkable. 
A variety of physics associated with the spontaneous breaking of 
continuous and discrete symmetries in certain fermionic models can 
now be studied with high precision using a meron-cluster algorithm.  
Questions related to chiral perturbation theory, physics of resonances, 
universality classes of phase transitions, etc. which appeared difficult
with hybrid Monte Carlo, can be studied much more easily. It is
important to extract as much as possible from these developments.

 In the context of finite density physics, there are many strongly 
interacting fermionic models that are still intractable. The example 
of QCD is a fundamental one. However, there are other models like the 
repulsive Hubbard model away from half filling or effective field 
theories of many nucleon systems, that are equally exciting to tackle. 
Although the solution to the problem in QCD still appears 
distant, the new progress suggests that solutions to the simpler
theories may not be too far away. However, one should to keep in
mind the difficulty of formulating cluster algorithms since  
the most interesting solutions to sign problems have emerged in the 
context of such algorithms. Limitations in cluster techniques 
may be correlated to our limitations in finding solutions to new sign 
problems.

 Based on the progress made so far many new questions for the 
future emerge. For example:
\begin{enumerate}
\item[(1)] Can one find a meron algorithm in a model whose ground state 
supports long range fermionic excitations? This can open up the
possibility of studying a variety of interesting quantum phase transitions.

\item[(2)] Is it possible to find an algorithm where one can update 
both fermionic and bosonic degrees of freedom together in an interacting 
model?

\item[(3)] Can one extend cluster algorithms to new types of spin 
and gauge models? This may be possible by understanding the constraints 
imposed by the topology of the configuration space \cite{Sok93}.
\end{enumerate}
The answers to these questions have the potential to determine if indeed 
the progress of the past couple of years is just the tip of an iceberg.

\vskip0.3in

\noindent {\bf Acknowledgement}

  I would like to thank my collaborators M. Alford, J. Cox and U.-J.Wiese 
for allowing me to present the preliminary results of the work on the 
Potts model. I would also like to thank the organizers of this conference 
for giving me the opportunity to present my work as a plenary talk. Finally 
I would like to thank the Institute for Nuclear Theory for its hospitality. 
Many ideas discussed here originated during the workshop on ``QCD at a finite
chemical potentials'' organized by the INT.


\begin{thebibliography}{9}
\bibitem{Kar00a} There is a factor of 3/2 between the normalization of $h$
used here and that used in \cite{Kar00}.
\end{thebibliography}
\end{document}